\documentclass[a4paper,10pt]{article}
\usepackage{amsfonts,amsthm,amsmath,amssymb,graphicx}

\newcommand{\be}{\begin{equation}}
\newcommand{\bea}{\begin{eqnarray}}
\newcommand{\ee}{\end{equation}}
\newcommand{\eea}{\end{eqnarray}}

\begin{document}

\makeatletter
\@addtoreset{equation}{section}
\makeatother
\renewcommand{\theequation}{\thesection.\arabic{equation}}

\rightline{WITS-MITP-009}
\vspace{1.8truecm}

\vspace{15pt}

%%%%%%%%%%%%%%%%%

{\large{
\centerline{   \bf Large trilinear $A_t$ soft supersymmetry breaking coupling from 5D MSSM}
\centerline {\bf  }
}}

\vskip.9cm

\thispagestyle{empty} \centerline{
     {\bf {Ammar Abdalgabar$^{{\dagger,\ddagger}}$}\footnote{ {\tt ammar.abdalgabar@students.wits.ac.za}} and A. S. Cornell$^{\dagger}$\footnote{ {\tt alan.cornell@wits.ac.za}}}
                                                    }

\vspace{.8cm}
\centerline{{\it $^{\dagger}$ National Institute for Theoretical Physics;}}
\centerline{{\it  School of Physics and Mandelstam Institute for Theoretical Physics,
}}
\centerline{{\it University of the Witwatersrand, Johannesburg.
Wits 2050, South Africa}}
\centerline{{\it $^{\ddagger}$ Department of Physics, Sudan University of Science and Technology,}}
\centerline{{\it Khartoum 407, Sudan }}

\vspace{1.4truecm}

%%%%%%%%%%%%%%%%%
\thispagestyle{empty}

\begin{abstract}
The possibility of generating a large trilinear $A_t$ soft supersymmetry breaking coupling at low energies through renormalisation group evolution in the 5D MSSM is investigated. Using the power law running in five dimensions and a compactification scale in the 10-$10^3$ TeV range, to show that gluino mass may drive a large enough $A_t$ to reproduce the measured Higgs mass and have a light stop superpartner below $\sim 1$ TeV as preferred by the fine tuning argument for the Higgs mass.
\end{abstract}

%%%%%%%%%%%%%%%%%%%%%%%%%%%%%%%%%%%%%%%%%%%%
%  Section 1: Introduction

\section{Introduction}

The observed scalar particle in 2012 of mass $\sim 125.5$ GeV \cite{ATLAS:2012zz, CMS:2012zz}, is consistent with the Standard Model (SM) Higgs boson. In the context of the Minimal Supersymmetric Standard Model (MSSM), this motivates considering models of supersymmetry breaking in which the stop superpartner is heavy (beyond the reach of the LHC) or a model in which a large trilinear $A_t$ soft supersymmetry breaking parameter can be generated at low energies. The first option is disfavored by fine-tuning arguments while the second
one allows for lighter stops $\sim 1$ TeV and it is thus preferred by naturalness arguments. The solution advocated in this work is to resolve the previous issue by introducing an extra (fifth) dimension and taking advantage of the power low running \cite{Abdalgabar:2013laa} to generate a sizable value of $A_t$ starting from a very small value.

\par We define 5-dimensional (5D) MSSM, the Higgs superfields and gauge superfields always live in the bulk. As consequence these fields will have Kaluza-Klein modes which contribute to the RGEs at $Q> 1/R$ and additional matter associated to five dimensional $\mathcal{N}=1$ super Yang-Mills. different possibilities of localising the matter superfields
can be studied. We shall consider the limiting case of superfields with SM matter fields restricted to the brane, and the RGEs for this scenario can be found in Ref. \cite{Abdalgabar:2014bfa}. Therefore there will be no additional Kaluza-Klein contributions of these matter fields to the RGEs\cite{Abdalgabar:2014bfa,Bhattacharyya:2010rm}.

\par We define 5-dimensional (5D) MSSM, the Higgs superfields and gauge superfields always live in the bulk. As consequence these fields will have Kaluza-Klein modes which contribute to the RGEs at $Q> 1/R$ and additional matter associated to five dimensional $\mathcal{N}=1$ super Yang-Mills \cite{Dienes:1998vh,Dienes:1998vg}. different possibilities of localising the matter superfields
can be studied. We shall consider the limiting case of superfields with SM matter fields restricted to the brane, and the RGEs for this scenario can be found in Ref. \cite{Abdalgabar:2014bfa}. Therefore there will be no additional Kaluza-Klein contributions of these matter fields to the RGEs\cite{Abdalgabar:2014bfa,Bhattacharyya:2010rm}.

\par Regarding the breaking of supersymmetry, whilst gauge mediation is favoured (and some recent work on gauge mediated supersymmetry breaking in a five dimensional context may be found in Ref. \cite{McGarrie:2013hca}), ultimately the universality of squark massses in GMSB means that even though the gaugino mediated limit \cite{Abdalgabar:2014bfa} might allow for light squarks (and 5D RGE evolution allows for a large $A_t$ and the observed Higgs mass), the collider bounds on first and second generation squarks \cite{Abdalgabar:2014bfa}, in the supra-TeV range would apply also to the $3^{rd}$ generation squarks, i.e. the stops, which as discussed before, is both phenomenologically less interesting and unnatural.  Therefore we wish for some other description of supersymmetry breaking that may allow for stops to be lighter than their first and second generation counterparts, such as in Refs. \cite{Brummer:2013upa,Abel:2014fka}. In this work we will therefore be unspecific about the precise details of how supersymmetry is broken and as a result also our conclusions will apply quite generally. We do however make some minimal specifications:
\begin{itemize}
\item We take as inputs the Yukawa and gauge couplings at the SUSY scale, $1$ TeV.
\item We will assume supersymmetry breaking occurs at the unification scale, which is found by finding the scale at which $g_1=g_2$, which is lowered compared to the 4D MSSM, by the effects of the compactification.
\item We specify the value of the gluino mass, $M_3$ at  $1$ TeV.
\item We take the trilinear soft breaking terms, $A_{u/d/e}$, to vanish at the unification scale.
\end{itemize}
We solved the combined set of differential equations numerically by using the above conditions, taking the ``third family'' approximation in which we only evolve third generation RGEs. This approximation is quite standard and is due to the relative smallness of the other Yukawa couplings (at least one order of magnitude) compared to those of the third generation and as a result the other A-term values are also very small.  We further specified some parameters such as $\mu$, $B_{\mu}$ and the value of the sfermion masses ($\sim 1$ TeV) so as to allow for the RGEs to be solved, but these do not affect the overall result.  We solved the differential equations between $Q_{min}=10^3$ GeV and $Q_{max}$, which was typically only one order larger than the unification scale.
%%%%%%%%%%%%%
\begin{figure}[h!]
\begin{center}
\includegraphics[width=7cm,angle=0]{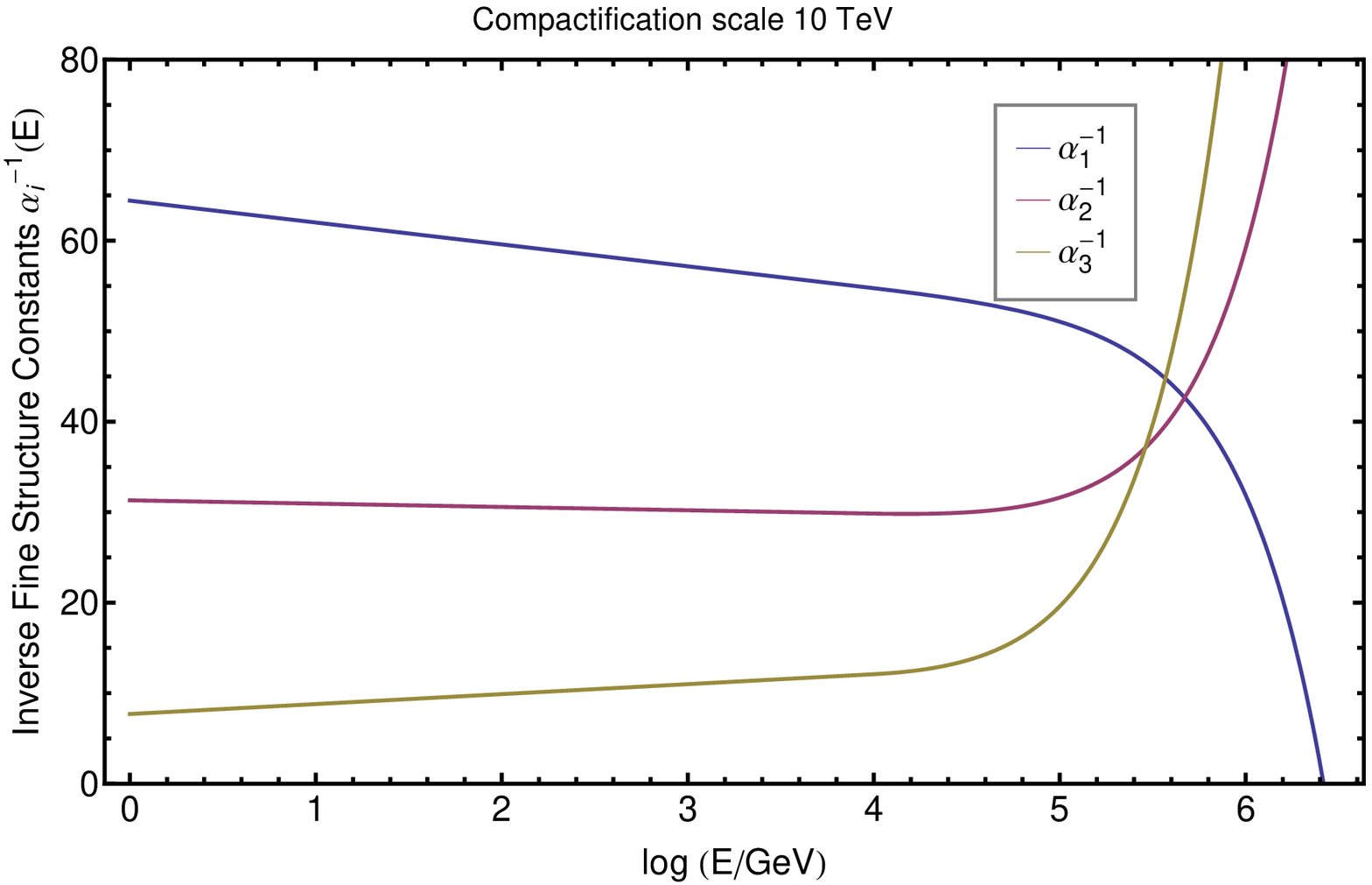}\qquad
\includegraphics[width=7cm,angle=0]{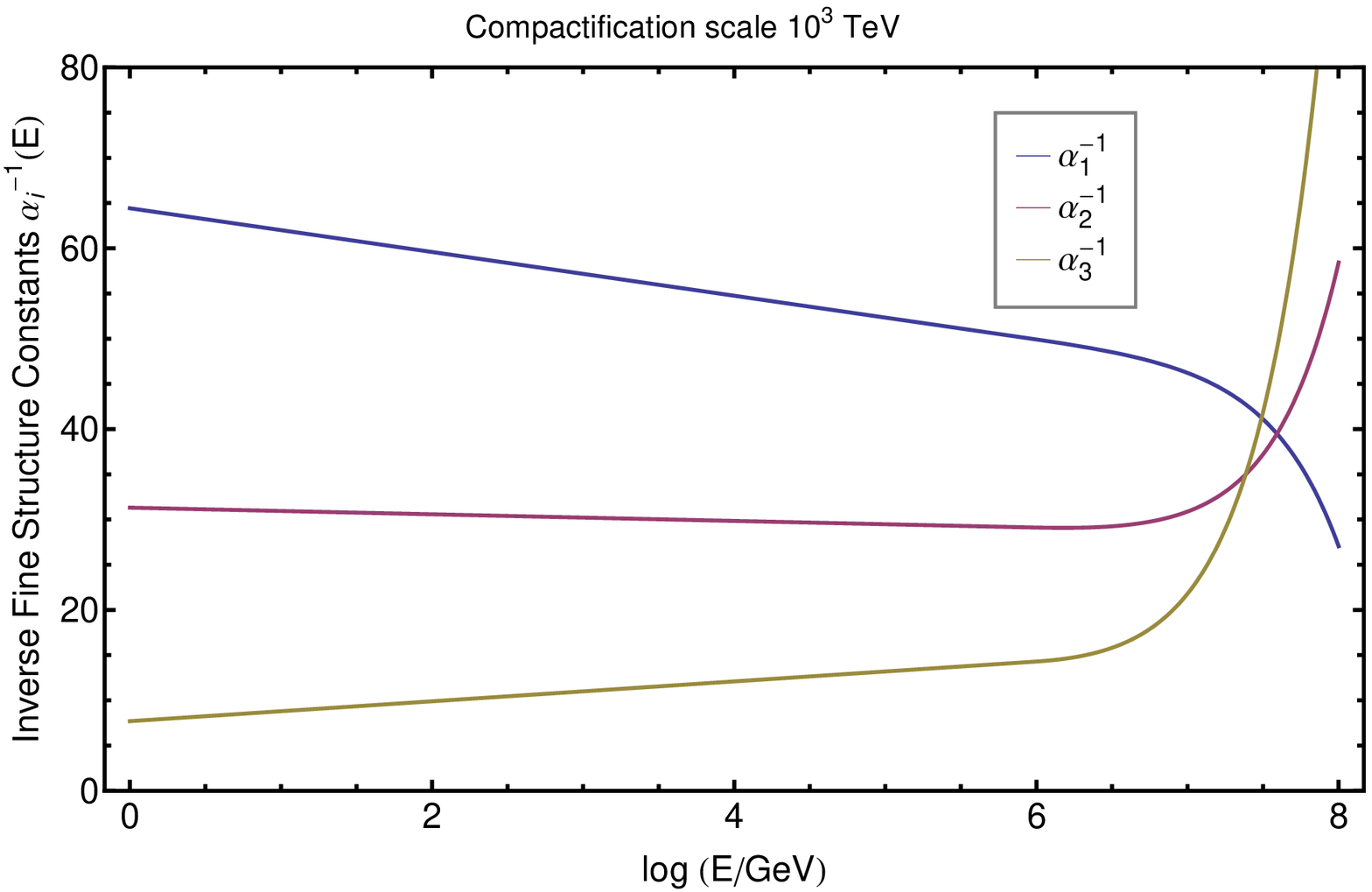}
\end{center}
\caption{{Evolution of the inverse fine structure constants $\alpha^{-1}(E)$, for two different values of the compactification scales 10 TeV (left panel), $10^3$ TeV (right), with $M_3$ of 1.7 TeV, as a function of  log(E/GeV).}}
\label{fig:alphas5D}
\end{figure}
%%%%%%%%%%%%%
%%%%%%%%%%%%%
\begin{figure}[h!]
\begin{center}
\includegraphics[width=7cm,angle=0]{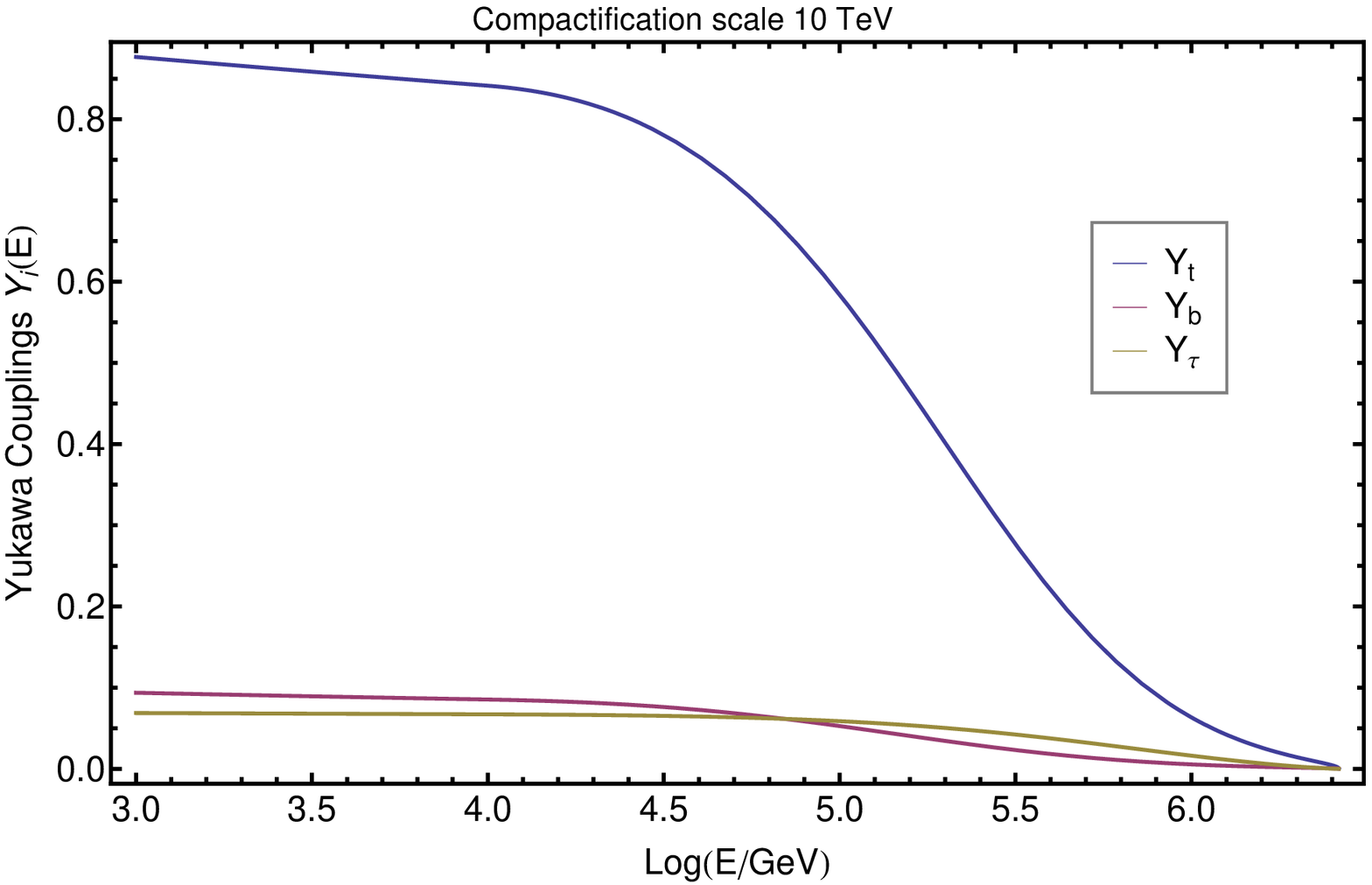}\qquad
\includegraphics[width=7cm,angle=0]{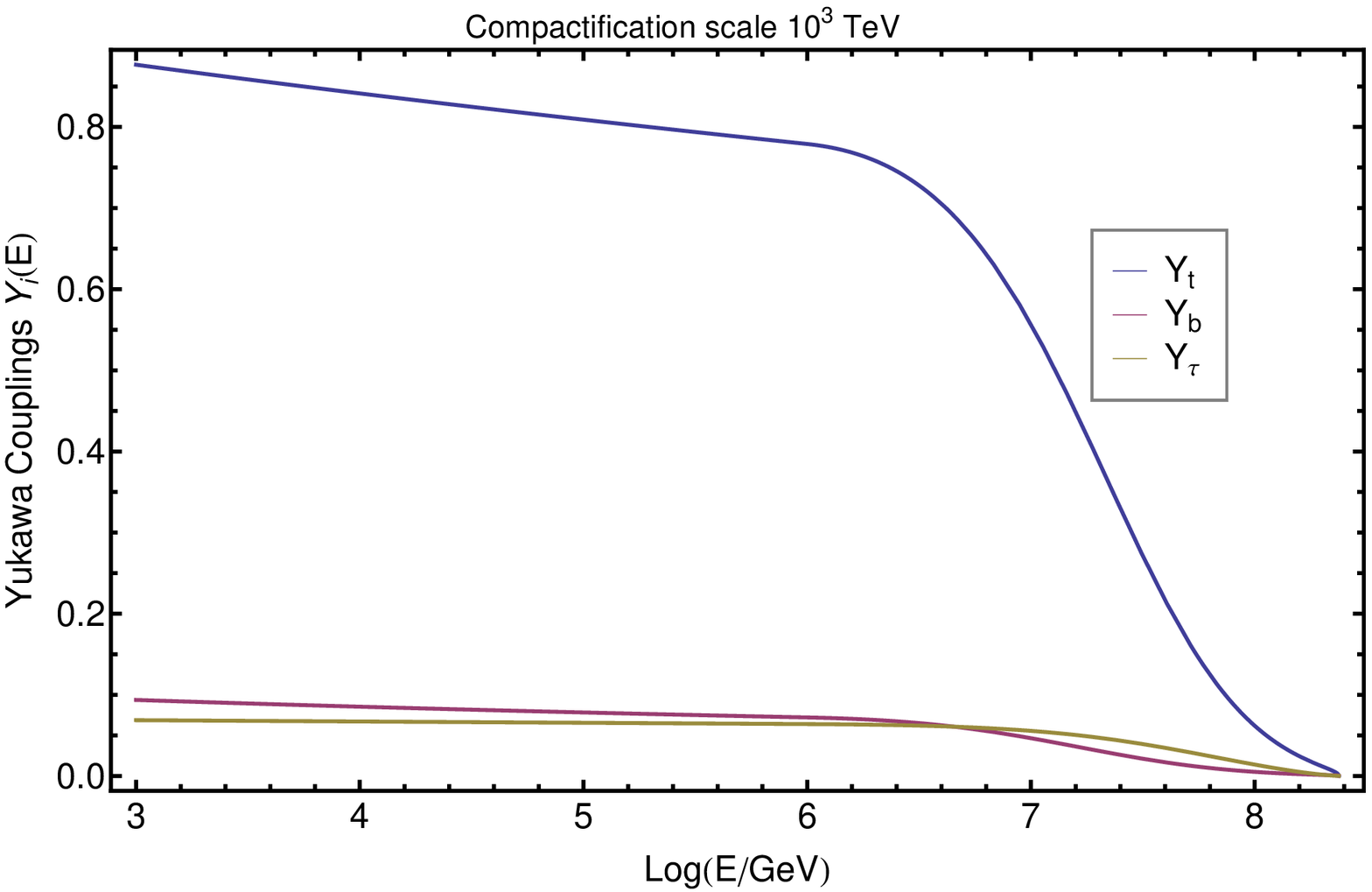}
\end{center}
\caption{{Evolution of Yukawa couplings $Y_i$,  for two different values of the compactification scales: 10 TeV (left panel), $10^3$ TeV (right), with $M_3[10^3]$ of 1.7 TeV, as a function of  log(E/GeV).}}
\label{fig:Yuk5D}
\end{figure}
%%%%%%%%%%%%%
%%%%%%%%%%%%%
\begin{figure}[h!]
\begin{center}
\includegraphics[width=7cm,angle=0]{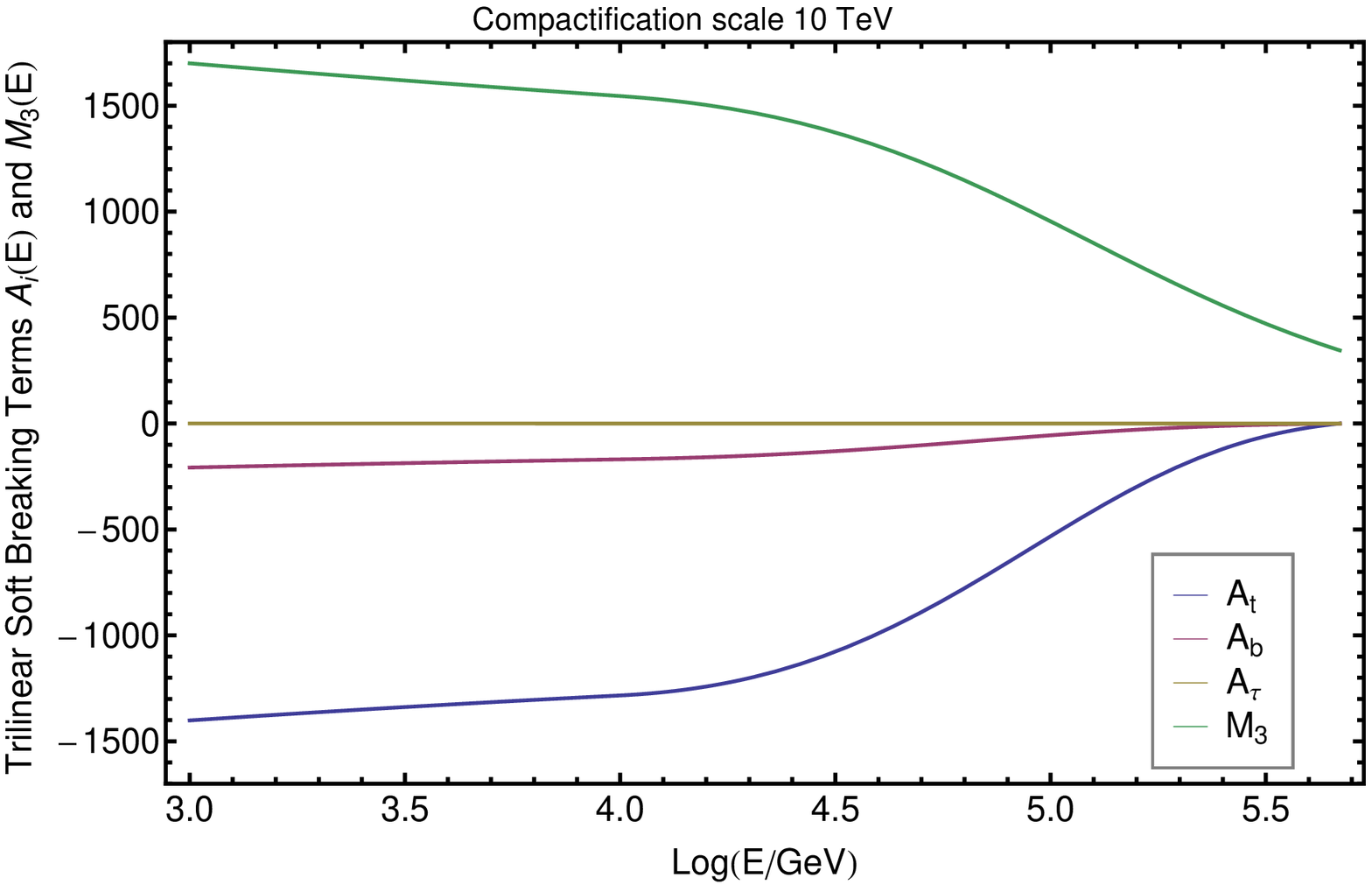}\qquad
\includegraphics[width=7cm,angle=0]{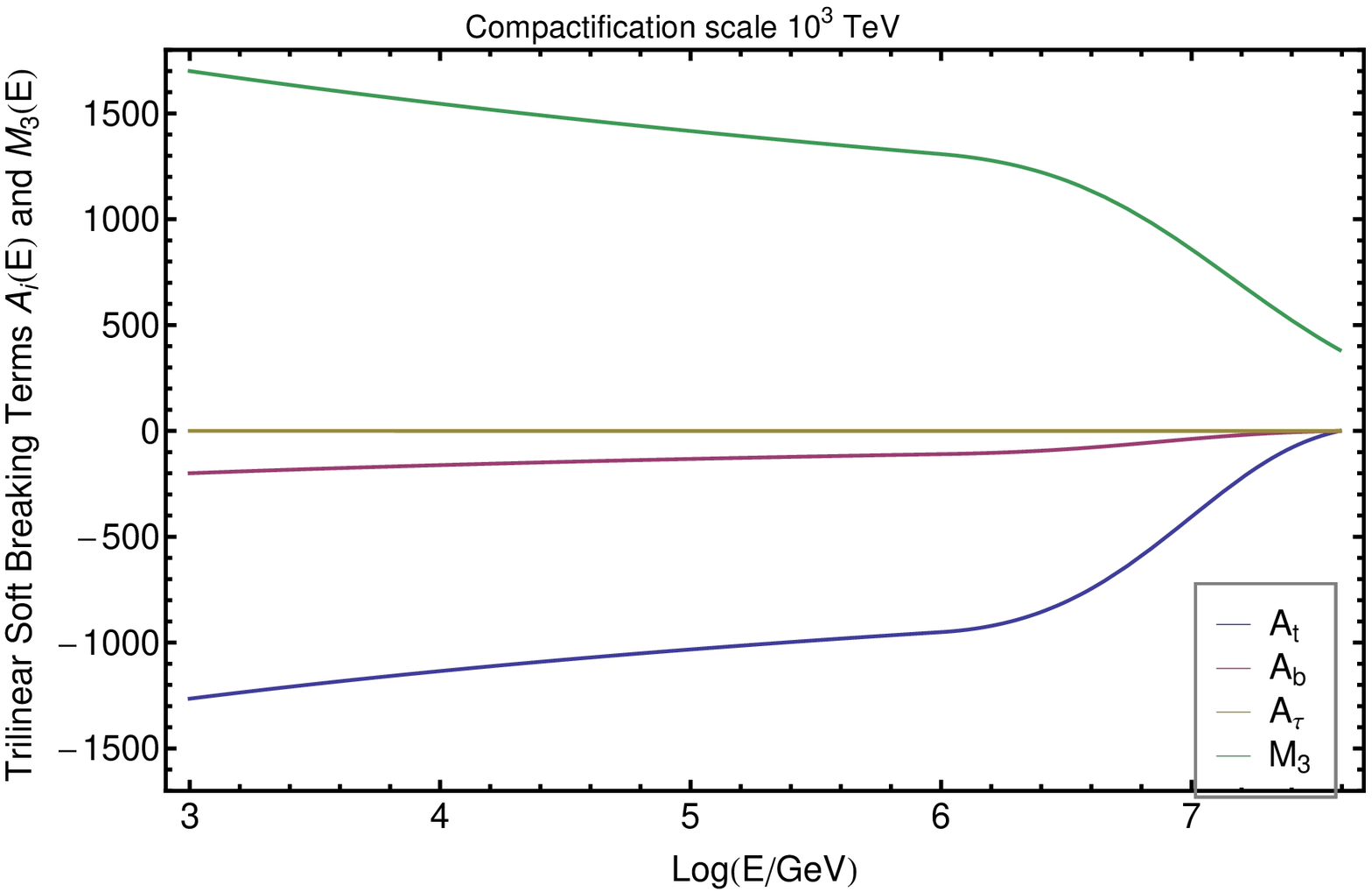}
\end{center}
\caption{{Evolution of trilinear soft terms $A_i(3,3)(E)$, for two different values of the compactification scales 10 TeV (left panel), $10^3$ TeV (right), with $M_3[10^3]$ of 1.7 TeV, as a function of log(E/GeV).}}
\label{fig:trilinearcompact}
\end{figure}
%%%%%%%%%%%%%
%%%%%%%%%%%%%
\begin{figure}[h!]
\begin{center}
\includegraphics[width=7cm,angle=0]{Trilinear10TeV1700M_3.eps}\qquad
\includegraphics[width=7cm,angle=0]{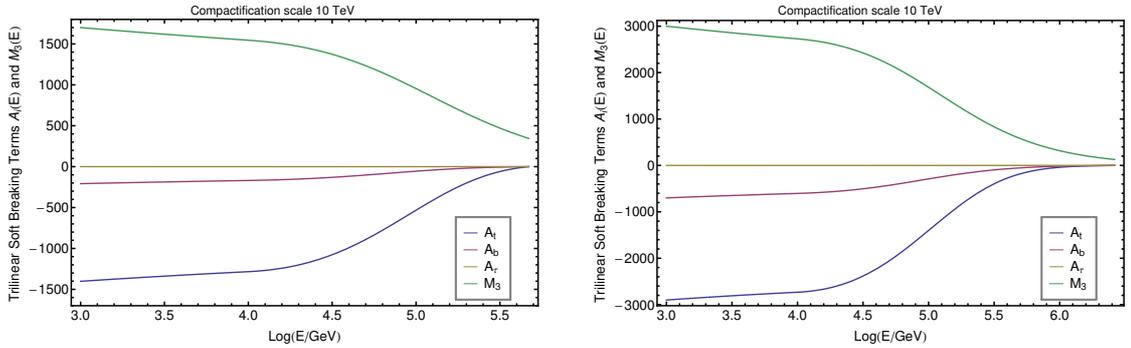}
\end{center}
\caption{{Evolution of trilinear soft terms $A_i(3,3)(E)$, for two different values of gluino masses, $M_3$: 1.7 TeV (left panel), 3 TeV (right panel), with $R^{-1}$ of 10 TeV, as a function of  log(E/GeV).}}
\label{fig:trilineargluino}
\end{figure}
%%%%%%%%%%%%%
%%%%%%%%%%%%%
\begin{figure}[h!]
\begin{center}
\includegraphics[width=0.9\textwidth]{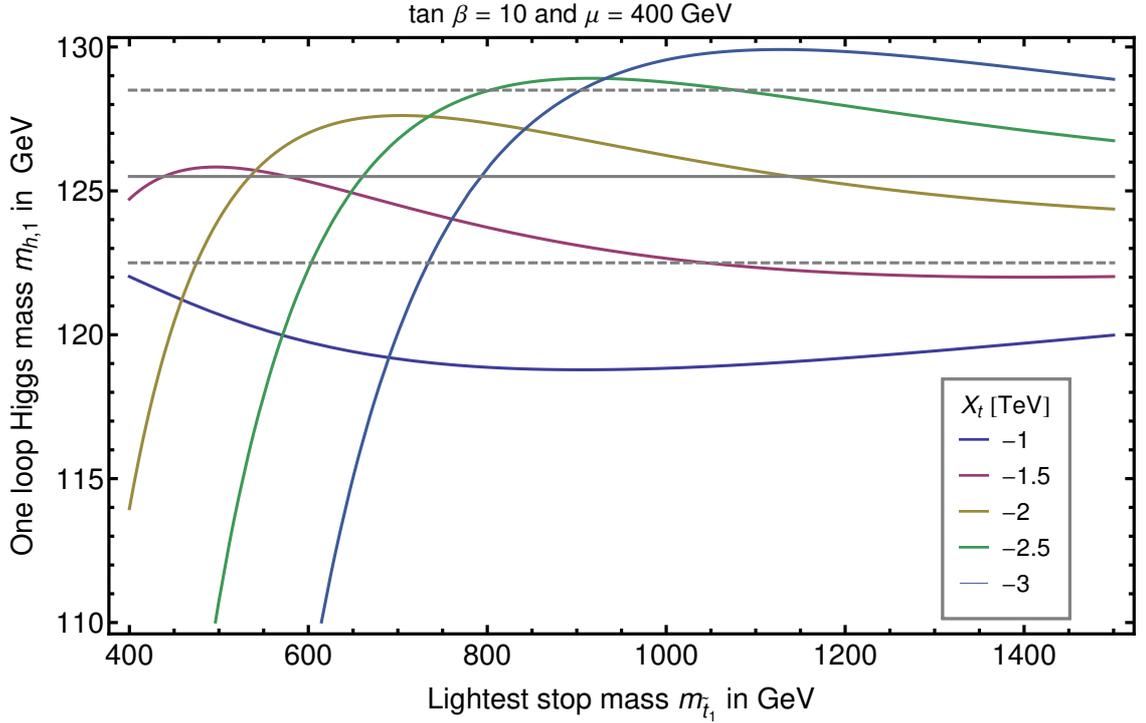}
\caption{{One-loop Higgs mass versus the lightest stop mass for representative values of $X_t=A_t-\mu \cot \beta$, corresponding to those of the 5D MSSM. }}
\label{fig:HiggsmassforAt}
\end{center}
\end{figure}
%%%%%%%%%%%%%
\begin{figure}[tph!]
\begin{center}
\includegraphics[width=7.5cm,angle=0]{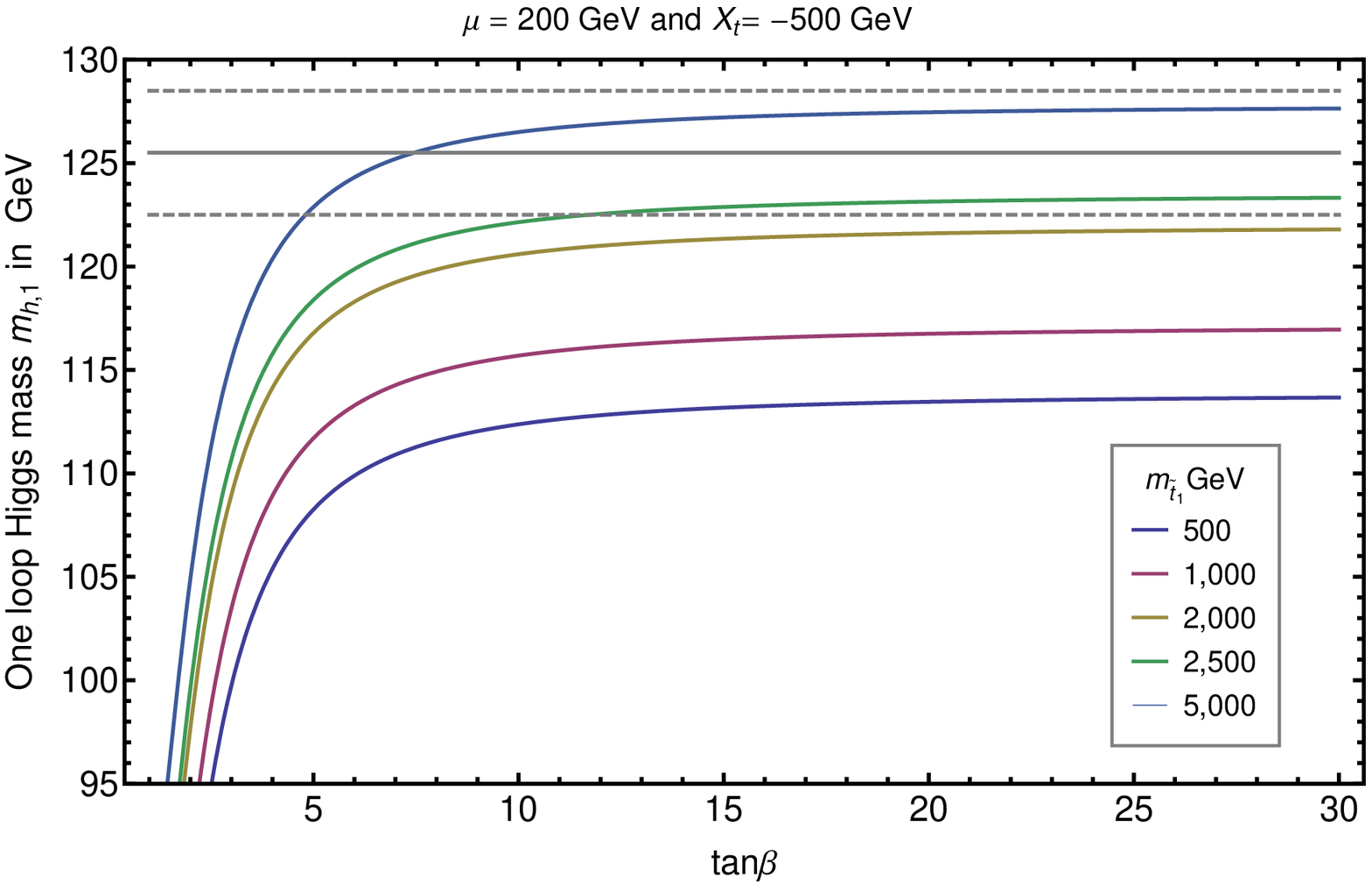}\qquad
\includegraphics[width=7.5cm,angle=0]{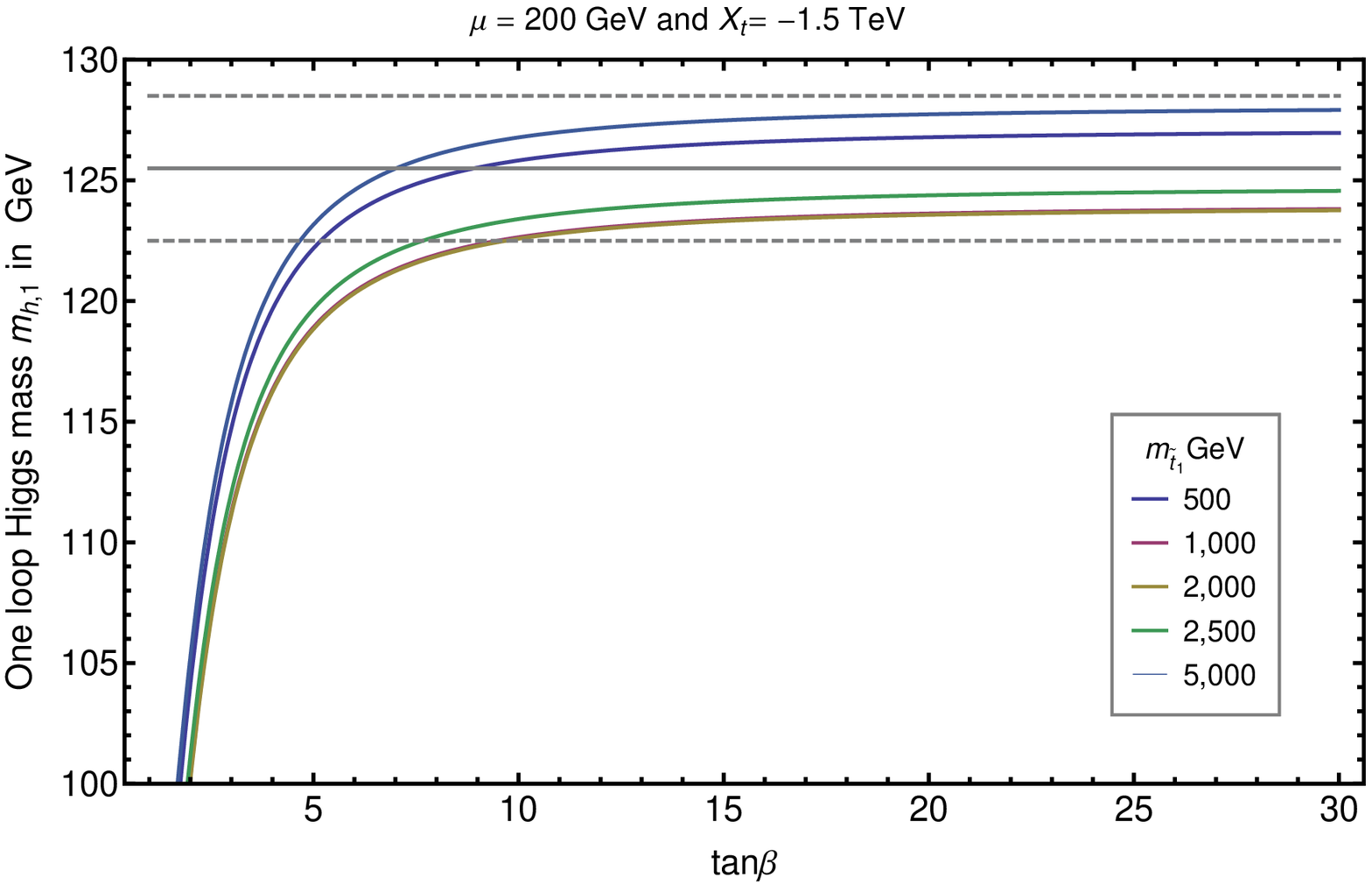}
\end{center}
\caption{{One-loop Higgs mass versus $\tan \beta$ for different values of the stop mass, for  $X_t=A_t-\mu \cot \beta$ of $-500$ GeV (left panel) and $-1.5$ TeV (right panel). }}
\label{fig:Higgsmassfortandbeta}
\end{figure}
%%%%%%%%%%%%%

\par A sufficiently large trilinear $A_t$ soft supersymmetry breaking parameter allows us to reproduce the measured Higgs field mass of $\sim 125.5$ GeV, while keeping a light stop superpartner (below $\sim 1$ TeV) as preferred by the fine tuning argument for the Higgs mass. Realising such a high $A_t$ is usually difficult (in supergravity, mGMSB, ...)\cite{Bharucha:2013ela,Brummer:2013upa,Abel:2014fka,McGarrie:2011av,Scrucca:2004cw,Falkowski:2005zv} but it is shown here concretely that 5D MSSM with compactification scale around 10-$10^3$ TeV can achieve large $A_t$ amounts at low scale (starting from $A_t(M_{GUT})\sim 0$), thanks to the power law running and simultaneously with an approximate unification of the gauge couplings, the precision of this gauge unification is illustrated quantitatively (at one-loop) for various values of the compactification scale as pictured in Fig. \ref{fig:alphas5D}. We also specify the Yukawa coupling RGE \cite{Abdalgabar:2014bfa} boundary conditions at $1$ TeV, which interestingly appears to vanish when evolved to the unification scale as shown in Fig. \ref{fig:Yuk5D}.

\par More precisely, it is found that increasing the compactification radius tends to increase the size of the trilinear $A_t$ parameter at low scale as shown in Fig. \ref{fig:trilinearcompact}. In fact the absolute value of $A_t$ mimics the magnitude of the final value of the gluino mass (at $1/R$) as can be seen in Fig. \ref{fig:trilineargluino}. As this interesting result of large $A_t$ generated radiatively is rather generic, the origin of supersymmetry breaking is left unspecified and the treatment adopted is effective.

\par This large $A_t$ can be realised in both cases of the three quark generations being on the boundary or only the third generation of matter fields on the boundary (``split families''). In the later case, the stop can be much lighter and possibly fall inside the LHC reach. All these results are based on the beta function for the gauge couplings, Yukawa couplings, trilinear terms, scalar soft masses, Higgs masses, $\mu$ and $B_{\mu}$ parameters, whose  analytical expressions can be found in Ref. \cite{Abdalgabar:2014bfa}.

\par Using the one-loop Higgs mass formula in Ref. \cite{Abdalgabar:2014bfa,Carena:1995bx,Haber:1996fp,Delgado:2004pr} and fixing $m_{h,1}=125.5$ GeV, $m_Z=91$ GeV, $\mu=200$ for $\tan \beta =10$ we can see in Fig. \ref{fig:HiggsmassforAt} that for representative values of $A_t$ achievable in the 5D MSSM, one may easily accommodate the lightest stop mass in the sub-TeV range. Let us pay attention to dependence on the value of $\tan \beta$ as pictured in Fig. \ref{fig:Higgsmassfortandbeta}.  In fact the value of $\tan \beta$ will depend greatly on how $\mu$ and $B_{\mu}$ are addressed in the context of supersymmetry breaking and hence the solution of the vacuum tadpole equations, but regardless of this, for values of $\tan \beta >10$ the functions are approximately flat and we expect the value to fall within this interval.  We expect that the $\mu$ term is naturally of the order of the electroweak scale, where in Fig. \ref{fig:HiggsmassforAt} we took a slightly large $\mu$ value of $400$ GeV and in Fig. \ref{fig:Higgsmassfortandbeta} we took $200$ GeV, leading typically to light Higgsinos and winos.

In conclusion, we have explored how 5D extension of the MSSM may generate large $A_t$ to achieve the observed Higgs mass and have sub-TeV stops, perhaps observable at the LHC.
We computed the full one-loop RGEs for all supersymmetric and soft breaking parameters. We find that the magnitude of $A_t$ follows closely that of the magnitude of $M_3$ and increases as the compactification scale decreases.

%%%%%%%%%%%%%%%%%%%%%%%%%%%%%%%%%%%%%%%%%%%%
\section*{Acknowledgments}

We would like to thank our collaborators Aldo Deandrea and Moritz McGarrie for their helpful discussions. This work is supported by the National Research Foundation (South Africa).

%%%%%%%%%%%%%%%%%%%%%%%%%%%%%%%%%%%%%%%
%  References
%

\end{document}